\documentclass{phc-proc4-auth}
\usepackage{amsmath}
\usepackage{amssymb}
\usepackage{graphicx}

\begin{document}

\begin{frontmatter}



\title{$d$-density wave state in an external
magnetic field}


\author[neuch]{S.G. Sharapov\corauthref{cor1}},
\author[kiev]{V.P. Gusynin},
\ead{vgusynin@bitp.kiev.ua}
\author[neuch]{H. Beck}
\ead{Hans.Beck@unine.ch}

\address[neuch]{Institut de Physique,
Universite de Neuch\^atel, Neuchatel, Switzerland}
\address[kiev]{Bogolyubov Institute for Theoretical Physics,
                03143 Kiev, Ukraine}
\corauth[cor1]{Corresponding author. Tel.: +41-32-718-3389;
fax: +41-32-718-2901; e-mail: Sergei.Sharapov@unine.ch}

\begin{abstract}
We present a study of the electrical and thermal conductivities of
the $d$-density wave (DDW) state in an external magnetic field $B$
in the low temperature regime and in the presence of impurities.
We show that in the zero temperature limit, $T \to 0$, the
Wiedemann-Franz (WF) law remains intact. For finite $T$ the WF law
violation is possible and it is enhanced by the external field.

\end{abstract}

\begin{keyword}
$d$-density wave \sep magnetic field \sep Wiedemann-Franz law
\PACS 71.10.-w \sep 74.25.Fy \sep 11.10.Wx \sep 74.72.-h

\end{keyword}
\end{frontmatter}

A recent experiment of Hill {\em et al.}  \cite{Hill:2001:Nature}
that measured electrical and thermal conductivities of the
optimally electron doped copper-oxide superconductor
Pr$_{2-x}$Ce$_x$CuO$_4$ (PCCO) in its normal state found striking
deviations from the Wiedemann-Franz (WF) law.  The hole
overdoped system Tl$_2$Ba$_2$CuO$_{6+\delta}$ (Tl-2201) was also studied
recently by Proust {\em et al.}
\cite{Proust:2002:PRL}. They verified that in the overdoped
Tl-2201 WF law holds perfectly.
The WF law is one of the basic properties of a Fermi liquid, reflecting
the fact that the ability of a quasiparticle to transport heat
is the same as its ability to transport charge, provided it cannot
lose energy through collisions. The WF law states that the ratio
of the heat conductivity $\kappa$ to the electrical conductivity
$\sigma$ of a metal is  a universal constant:
\begin{equation}
L_0 \equiv \frac{\kappa}{\sigma T} = \frac{\pi^2}{3}
\left( \frac{k_B}{e}\right)^2,
\end{equation}
where $k_B$ is the Boltzmann's constant, $e$ is electron's charge
and $L_0 = 2.45 \times 10^{-8} \mbox{W} \Omega \mbox{K}^{-2}$ is
Sommerfeld's value for the Lorenz ratio $L \equiv  \kappa/(\sigma
T)$. To be more precise, one should also specify the temperature
range where the WF law holds. Strictly speaking this law is proven
only in the limit $T \to 0$ and for a small concentration of
impurities \cite{Langer:1962:PR}.
One of the possible theoretical interpretations of the WF law breakdown is that the
quasiparticle fractionalizes into separate spin and charge.
This separation can be investigated using
various models and approaches.

An examination of the WF law was done by Yang and Nayak (YN)
\cite{Yang:2002:PRB} and also by Kim and Carbotte (KC) \cite{Kim:2002:PRB}
within the phenomenological $d$-density wave (DDW) picture.
The DDW scenario proposed in Ref.~\cite{Chakravarty:2001:PRB}
is based on the assumption that the pseudogap
phenomenon \cite{Timusk:1999:RPP} in high-$T_c$ cuprates is the result
of the development of another order parameter called DDW order
that has $d$-wave symmetry and can be described by the mean-field
Hamiltonian
\begin{equation}
\label{Hamiltonian.DDW.matrix}
H^{\mathrm{DDW}}  = \int_{\mathrm{RBZ}} \frac{d^2 k}{(2 \pi)^2}
\chi_{s}^{\dagger}(t, \mathbf{k}) \left[ H_0(\mathbf{k}) - \mu
\right] \chi_{s}(t, \mathbf{k}),
\end{equation}
where
\begin{equation}
\label{H}
H_0(\mathbf{k}) =\varepsilon(\mathbf{k}) \sigma_3 - D(\mathbf{k}) \sigma_2,
\end{equation}
the spinors $\chi_s$ and $\chi_s^{\dagger}$ are
\begin{equation}
\label{Nambu.variables}
\chi_s^{\dagger}(t, \mathbf{k}) = \left( \begin{array}{cc}
c_{s}^{\dagger}(t, \mathbf{k})
\quad c_{s}^{\dagger}(t, \mathbf{k + Q})
\end{array} \right),
\end{equation}
the single particle energy is $\varepsilon({\mathbf{k}}) = -
2 t (\cos k_x a + \cos k_y a)$
with $t$ being the hopping parameter,
$\mu$ is the chemical potential,
$D(\mathbf{k}) = \frac{D_0}{2}(\cos k_x a - \cos k_y a )$ is the
$d$-density wave gap and $\mathbf{Q} = (\pi/a, \pi/ a)$ is the
wave vector at which the density-wave ordering takes place
and $\sigma_i$  are Pauli matrices. The
integral is over the reduced Brillouin zone. The units
$\hbar = k_{B} = c= 1$ are chosen.

One of the unusual features of the DDW state is that for a half-filled band
the chemical potential of the nodal quasiparticles participating in the electrical
and thermal transport can be small or even zero, i.e. $|\mu| < k_B T$,
that violates the usual conditions of the WF law validity.
Indeed, exactly in the limit $\mu =0$ the WF law is strongly violated in
the extremely clean
limit \cite{Yang:2002:PRB}.
There is no WF law violation in the $T =0$ limit for finite $\mu$ and/or $\Gamma$
\cite{Yang:2002:PRB,Kim:2002:PRB}. For finite
temperatures the WF violation depends on the impurity scattering \cite{Kim:2002:PRB}:
in the Born limit (for a constant impurity scattering rate $\Gamma$)
there is no change in the WF law, but in the unitary limit for the frequency
dependent scattering rate the WF law is violated, but only for $|\mu|$
smaller that the DDW gap. When $\mu$ is increased sufficiently, the Lorenz number
becomes approximately equal to its conventional value
and its temperature dependence is small.

While in general the validity of the DDW pseudogap scenario is
still questionable, it is important to scrutinize all its
theoretical consequences. One of the opportunities is to study
possible WF law violations using the DDW model, so that these
results can be compared with the experimental results of
Refs.~\cite{Hill:2001:Nature,Proust:2002:PRL}. {\em The presence
of the external magnetic field\/}  is an essential ingredient of
the experiments \cite{Hill:2001:Nature,Proust:2002:PRL}, so that
if the DDW state exists, it would show up in the magnetic field in
the underdoped regime at low $T$ when the superconductivity is
destroyed. In this paper we present the study
\cite{Sharapov:2003:PRB} of the WF law for the DDW model {\em
including\/} the external magnetic field for a constant impurity
scattering rate paying special attention to the regime $|\mu|
\lesssim k_B T$ where the violation of the WF law is expected.

\begin{figure}[h!]
\centering{
\includegraphics[width=6.5cm]{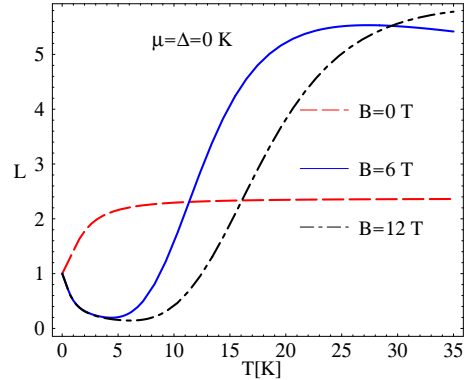}}
\caption{The normalized Lorenz number $L/L_0$
as function of temperature, $T$, for three different values of magnetic
field $B$ at half-filling, $\mu =0$.}
\label{fig:1}
\end{figure}

We show that the strongest violation of the WF law is possible for $\mu = 0$
and only at finite temperatures. This is the case shown in Fig.~\ref{fig:1}.
We observe that while the results of \cite{Hill:2001:Nature}
suggest that the WF law is violated at $T \to 0$, {\it there is
no} violation of the WF law in this limit in the DDW scenario of
the pseudogap.   Since in Fig.~3 of Ref.~\cite{Hill:2001:Nature}
the electrical conductivity is a constant, the line
$\kappa_e(T)/T$ directly represents the normalized Lorenz number
$L(T)/L_0$. For finite temperatures there is then some similarity
between Fig.~\ref{fig:1} (see also other figures in
Ref.~\cite{Sharapov:2003:PRB}) and Fig.~3 of
\cite{Hill:2001:Nature} where as $T$ increases the thermal
conductivity crosses from the region with $\kappa_e/T <
L_0/\rho_0$ to the region with $\kappa_e(T)/T > L_0/\rho_0 $
resembling the character of the WF law violation seen in the
experiment. It is obvious from Fig.~\ref{fig:1} that such a
behavior of $L(T)$ is due to the {\it presence} of the magnetic
field. This confirms our claim that to interpret theoretically the
experiment \cite{Hill:2001:Nature} one should take into account
the influence of the external field.

Our main conclusions concerning the WF law
can be summarized as follows.

\noindent
(1) We have shown that
in the DDW state in the presence of impurities
the WF law holds in $T \to 0$ limit for an arbitrary field $B$
and chemical potential $\mu$.
This is checked within the bubble approximation, i.e. not including
the impurity vertex.

\noindent
(2) For finite temperatures $T \lesssim |\mu|$, the WF law violation
is possible and in zero field the thermal conductivity dominates over
the electrical conductivity, i.e. $L(T)/L_0 > 1$.

\noindent
(3) For  $T \lesssim |\mu|$
in the nonzero field the WF law violation becomes even
stronger than in zero field and depending on the temperature both regimes
 $L(T)/L_0 \ll 1$  and $L(T)/L_0 > 1$ are possible.

\noindent
(4) For $T \ll |\mu|$ there is no WF violation even in the presence of
magnetic field.

This work was supported by the research project
20-65045.01 of the Swiss NSF. The work of V.P.G. was supported by
the SCOPES-projects 7UKPJ062150.00/1 and 7 IP 062607 of the Swiss
NSF and by the Grant No. PHY-0070986 of NSF (USA).

\end{document}